\documentclass[a4paper]{jpconf}
\usepackage{graphicx}
\begin{document}
\title{A cavity-mediated collective quantum effect \\ in sonoluminescing bubbles}

\author{Almut Beige and Oleg Kim}

\address{The School of Physics and Astronomy, University of Leeds, Leeds LS2 9JT, United Kingdom}

\ead{a.beige@leeds.ac.uk}

\begin{abstract}
This paper discusses a collective quantum effect which might play an important role in sonoluminescence experiments. We suggest that it occurs during the final stages of the collapse phase and enhances the heating of the particles inside the bubble.
\end{abstract}

\section{Introduction}

\begin{figure}[b]
\begin{center}
\includegraphics[width=110mm]{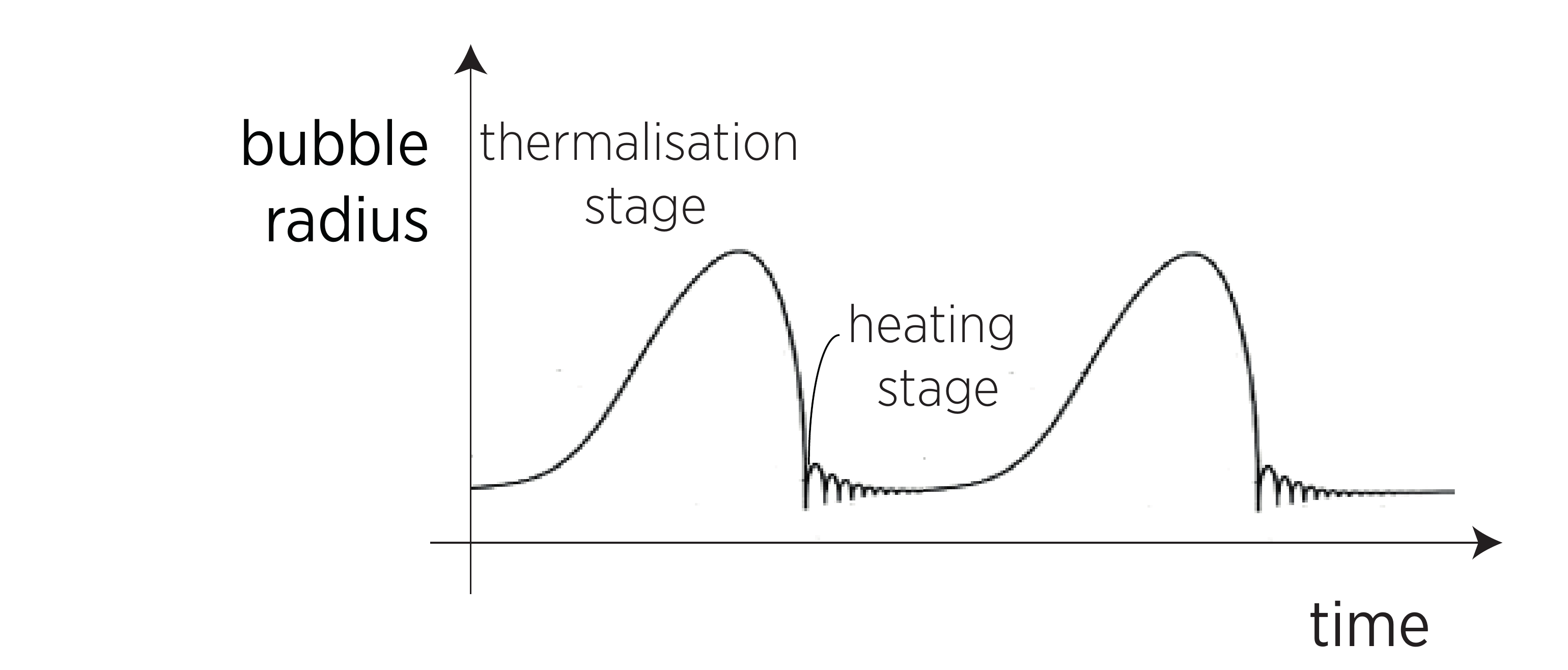}
\end{center}
\caption{\label{label} Schematic view of the time dependence of the bubble radius in a typical single-bubble sonoluminescence experiment. Most of the time, the bubble radius increases isothermal (thermalisation stage). During the collapse phase (heating stage), the temperature within the bubble increases rapidly and a strong light flash occurs.}
\end{figure}

Sonoluminescence is the intriguing phenomenon of strong light flashes from tiny bubbles in a liquid \cite{Brenner}. The bubbles are driven by an ultrasonic wave and need to be filled with atomic species. Fig.~\ref{label} shows a typical single-bubble sonoluminescence cycle \cite{Brenner,Lohse}. For most of the cycle, the bubble radius increases isothermal and is in good agreement with the laws of classical physics \cite{Moss2}. However, each expansion phase is followed by a very rapid collapse phase during which the bubble becomes thermally isolated from the liquid. Close to its minimum radius of about $0.5 \, \mu$m, a rapid heating of the particles inside the bubble occurs which is accompanied by the sudden emission of light. Afterwards a re-expansion phase begins in which the bubble oscillates around its equilibrium radius until it regains its stability. 

Measuring the spectrum of the picosecond light flash and associating it with blackbody or {\em Bremsstrahlung} radiation indicates temperatures of at least $10^3$--$10^4 \,$K inside the bubble \cite{Barber1,Barber3,Vazquez,Vazquez2,Suslick5,McNamara2}. It is even possible to observe light emission in the ultraviolet regime which hints at temperatures of about $10^6 \,$K \cite{Camara}. Noteworthy is the discovery of sharp emission lines in the optical regime \cite{Brenner,Suslick5,Suslick3}. These indicate the population of highly excited energy eigenstates of noble gas and metal atoms which cannot be populated thermally but hint at the presence of a dense plasma \cite{Suslick5,Suslick3,Suslick1,Suslick2,Suslick4,Flannigan2,Gordeychuk}.

Although sonoluminescence has been studied extensively, the origin of the sudden energy concentration during the final stage of the bubble collapse phase remains a mystery \cite{Suslick5,Putterman2}. A valid theoretical model needs to be based on the formation of a plasma and needs to include a mechanism which can increase the temperature of this plasma by at least one order of magnitude. It needs to be able to operate in a solid state-like environment and on very short time scales. Moreover, the proposed heating mechanism needs to depend strongly on the atomic species inside the cavitating bubble. For example, single bubble sonoluminescence experiments with ionic liquids reach significantly higher temperatures than experiments based on noble gas atoms. The purpose of this paper is to emphasise that such a heating mechanism could be based on a recently-identified cavity-mediated collective quantum effect \cite{kim2}. More details of its relevance in sonoluminescence experiments can be found in Refs.~\cite{sono2,sono3}.

More concretely, we assume in the following that the electromagnetic field inside the bubble becomes quantised when the bubble approaches its minimum radius. During the collapse phase, the atomic particles therefore behave as if placed inside a so-called optical resonator. The walls of the cavity are formed by the walls of the collapsing bubble. As in Ref.~\cite{sono}, we moreover quantise the motion of the atomic species inside the bubble at this stage. The atoms are now so strongly-confined that they cannot be compressed any further. Simultaneously, the collapse of the bubble results into high enough temperatures to lead to the formation of a plasma. Once the plasma is formed, collisions create correlations between the electronic and the vibrational states of the atomic particles which can fuel a cavity-mediated collective quantum optical heating process \cite{sono2,sono3}. As in ion trap experiments \cite{cooling} and in cavity-mediated laser cooling \cite{kim2,kim}, the rapid temperature changes of the confined particles are due a complex interplay between electronic and vibrational degrees of freedom combined with the spontaneous emission of photons. If repeated over many cycles, such a mechanism can have a significant effect. 

\section{A collective effect in a many-body quantum system} \label{toymodel}

Before describing its relevance to sonoluminescence experiments, this section introduces a highly-simplified quantum optical toy-model. Our aim is to illustrate a collective effect in the dynamics of a many-body quantum system which changes some properties of the interacting particles very rapidly. The model is the following: Suppose $N$ particles have two different degrees of freedom which can be described by a two sets of bosonic operators $a_i$ and $b_i$ with $i=1,...,N$. Moreover, the particles couple to a common mode which can be modelled by another bosonic operator $c$. Then
\begin{eqnarray} 
\left[ a_i , a_i^\dagger \right]  = \left[ b_i , b_i^\dagger \right] = \left[ c, c^\dagger \right] &=& 1 \, . 
\end{eqnarray}
In addition, suppose there is a direct coupling between the internal degrees $a$ and $b$ of the particles and the common mode $c$ which can be modelled by an interaction Hamiltonian of the form
\begin{eqnarray} \label{HI2first}
H_{\rm I} &=& \sum_{i=1}^N \hbar \eta g \, \left(b_i + b_i^\dagger \right) \left( a_i c^\dagger + a_i^\dagger c \right) \, .
\end{eqnarray}
Here $\eta g$ is a three-body interaction constant which depends on the physical properties of the modelled quantum system. In the following, we use this Hamiltonian to analyse the dynamics of the mean average excitation number $m$ of the $b$-mode given by
\begin{eqnarray} \label{notation}
m \equiv {1 \over N} \sum_{i=1}^N \left \langle b_i^\dagger b_i \right \rangle \, .
\end{eqnarray}
To get a closed set of rate equations, we moreover need to consider the expectation values 
\begin{eqnarray} \label{notation2}
k_1 &\equiv & {1 \over N} \sum_{i=1}^N \left \langle {\rm i} \left( b_i - b_i^\dagger \right) \left( a_i c^\dagger + a_i^\dagger c \right) \right \rangle \, , \nonumber \\
k_2 &\equiv & {1 \over N(N-1)} \sum_{i=1}^N \sum_{j \neq i} \left \langle a_i b_i^\dagger a_j^\dagger b_j - a_i b_i a_j^\dagger b_j^\dagger \right \rangle \, .
\end{eqnarray}
All three expectation values are real and normalised such that they remain essentially the same in the large $N$ limit, if the number of particles in the trap changes.  

Now we can use Schr\"odingers equation, which implies $\langle \dot A_{\rm I} \rangle = - ({\rm i}/\hbar) \langle [A_{\rm I}, H_{\rm I}] \rangle$ for the time evolution of the expectation value of any physical observable $A_{\rm I}$ in the interaction picture, to analyse the dynamics of the above introduced variables. Doing so, we find that 
\begin{eqnarray} \label{notation3}
\dot m \, = \, \eta g \, k_1 ~~ {\rm and} ~~ \dot k_1 \, = \, - 2 N \eta g \, k_2 
\end{eqnarray}
to a very good approximation. In the time derivative of $k_1$, only terms which scale as the number $N$ of particles involved have been taken into account. If $k_2$ differs from zero and $N \gg 1$, the above differential equations imply that the mean excitation number $m$ of the $b$-mode changes very rapidly in time, namely at a rate which scales as $N$. In other words, we identified a collective effect in the dynamics of a many-body quantum system with three-body interactions.

\section{Application of the above collective quantum effect to sonoluminescing bubbles}

 As shown in Refs.~\cite{sono2,sono3}, when modelling the collapse phase of a sonoluminescing bubble, we use operators analogous to $a$ and $b$ to describe the electronic and the vibrational degrees of freedom of the confined atomic particles, while $c$ refers to the quantised electromagnetic field inside the bubble. When the atoms accumulate in the centre of the bubble during the final stages of the collapse phase, their dynamics is governed by an interaction Hamiltonian which is essentially the same as $H_{\rm I}$ in Eq.~(\ref{HI2first}). Consequently, the mean number of excitations in the $b$-mode can change on a time scale which scales as the number $N$ of atomic particles inside the bubble. The result is a collective change of the mean number of phonons in the system, ie.~very rapid heating or cooling. This process is fuelled by a non-zero expectation value $k_2$. Here $k_2$ is expected to arise during the thermalisation phase due to atomic collisions which result in the exchange of electronic and vibrational energy between the particles \cite{sono2,sono3}. If this process is repeated over many cycles, the described mechanism can have a significant effect on the temperature of the bubble. 

Combined with the ability of optical cavities to spontaneously emit photons and the similarities of the sonoluminescing bubble with the experimental setups used for laser cooling \cite{sono,cooling} and cavity-mediated laser cooling \cite{kim2,kim}, the above described interaction suggests that the atomic particles can change their temperature very rapidly, once the cavity approaches its minimum radius. To obtain a more intuitive picture and to show that the result is indeed heating and not cooling, notice that the frequency of the cavity field is relatively small, as long as the bubble radius is relatively large. Even when the bubble reaches its minimum radius, its frequency is always (first) below the transition frequency of the confined atomic particles. Energy conservation therefore implies that the conversion of atomic excitation into cavity photons is in general accompanied by the creation of a phonon. When the cavity photon leaks out of the resonator, thereby contributing to the strong light flash seen at the end of the collapse phase, the simultaneously created phonon remains inside the bubble. Subsequently, more phonons are created after the atoms have been re-excited during another thermalisation stage.

Finally, we remark that the strength of the above described heating mechanism depends on the physical properties of the sonoluminescing bubble, like the size of the relevant atomic transition frequency with respect to the minimum bubble radius. For noble gas atoms, the atomic transition frequency is in general very high so that the above mentioned conversion of atomic excitation into cavity photons and phonons is in general highly detuned and relatively inefficient. However, in ionic liquids, such conversions become much more likely due to being almost in resonance.

\section{Conclusions}

This paper discusses a collective quantum effect which might play an important role during the final stages of the collapse phase of sonoluminescence experiments. Although the calculations in Section \ref{toymodel} only provide a very rough picture of the dynamics of a sonoluminescing bubble, it captures the origin of a cavity-mediated collective quantum optical heating mechanism which we identify in Ref.~\cite{sono2,sono3}. \\[0.3cm]

\ack AB and OK thank the UK Engineering and Physical Sciences Re- search Council EPSRC for financial support (Grant Ref.~EP/H048901/1). Moreover, we acknowledge stimulating discussions with Antonio Capolupo, Andreas Kurcz, and Daniel Rumgay.

\end{document}